\documentclass{appolb}
\usepackage{cite}
\usepackage{epsfig}
\usepackage{amsfonts,amsmath}
\usepackage{hhline}
\usepackage{amssymb}
\usepackage{times}
\usepackage{cite}

\newcommand{\kperp}{k_\perp}
\newcommand{\alps}{\alpha_s}
\newcommand{\ordalps}{{\cal O}(\alpha_s)}

\newcommand{\Xbj}{x_{\rm Bj}}

\begin{document}
\title{Three-Jet Production in Deep-Inelastic Scattering at HERA
\thanks{Presented at DIS2002, April 30 - May 4, Cracow, Poland}%
}
\author{Christian Schwanenberger, DESY
\address{on behalf of the H1 Collaboration}
% {Deutsches Elektronen-Synchrotron DESY, Notkestra{\ss}e 85,\\ 
%D-22607 Hamburg, Germany\\ 
%E-mail: schwanen@mail.desy.de}
%\and
%on behalf of the H1 Collaboration
}
\maketitle
\begin{abstract}
Three-jet production has been studied in 
deep-inelastic positron-proton scattering.
The measurement carried out with the H1 detector at HERA
covers a large range of four-momentum transfer squared
$5 < Q^2 < 5\,000\,{\rm GeV}^2$ and invariant three-jet masses
$25 < M_{\rm 3jet} \lesssim 140\,{\rm GeV}$.
Jets are defined by the inclusive $\kperp$ algorithm 
in the Breit frame.
The size of the three-jet cross section and the ratio of 
the three-jet to the dijet cross section $R_{3/2}$ are 
described over the whole phase space by the predictions 
of perturbative QCD in next-to-leading order.
The shapes of angular jet distributions deviate significantly 
from a uniform population of the available phase space
but are well described by the QCD calculation.
\end{abstract}

% **********************************************************************
% **********************************************************************
% **********************************************************************
\section{Introduction}\label{sec:intro}

Multi-jet production in deep-inelastic scattering (DIS)
has been successfully used at HERA to test the predictions of 
perturbative QCD (pQCD) over a large range
of four-momentum transfer squared $Q^2$. 
Recently the H1 collaboration has determined the strong coupling 
$\alps$ and the gluon density in the proton~\cite{h1gluon}
from the inclusive jet and the dijet cross sections measured 
in the Breit frame.
While these cross sections are directly sensitive to QCD effects 
of $\ordalps$, the three-jet cross section in DIS is already 
proportional to $\alpha_s^2$ in leading order (LO) in pQCD.
The higher sensitivity to $\alpha_s$ and the greater number of
degrees of freedom of the three-jet final state thus
allow the QCD predictions to be tested in more detail.
Here we present differential 
measurements of the three-jet cross section in 
neutral current DIS and measurements of shapes of angular 
jet distributions which are sensitive to dynamic effects 
of the interaction~\cite{bib:paper}.
The present analysis includes the first comparison of three-jet 
distributions measured in hadron induced reactions with a 
pQCD calculation in next-to-leading order (NLO) i.e. {${\cal
    O}(\alpha_s^3)$}~\cite{nlojet}.

\section{Measured Observables}

In neutral current DIS the lepton interacts with a parton 
in the proton via the exchange of a boson ($\gamma$, $Z$).
At a fixed center-of-mass energy the kinematics of the 
lepton inclusive reaction (for unpolarized lepton and proton beams)
is given by two variables
which are here chosen to be the four-momentum transfer squared
$Q^2$ and the Bjorken scaling variable $\Xbj$.
The subprocess   $1 + 2 \rightarrow  3 + 4 + 5$ 
in which three massless jets emerge from the boson-parton reaction 
is fully described by six further variables
which can be constructed from the energies $E_i$
and the momenta $\vec{p}_i$ of the jets in the 
three-jet center-of-mass (CM) frame.
It is convenient to label 
the three jets ($i = 3, 4, 5$) in the order
of decreasing energies in the three-jet CM frame.
Two of these variables are the angles $\theta_3$ and $\psi_3$ that specify
the relative orientation of the jets,
\begin{equation}
\cos{\theta_3} \equiv \frac{\vec{p}_{B} \cdot \vec{p}_3}
                     { |\vec{p}_{B}| \; | \vec{p}_3 |}     \, ,
  \hskip11mm 
\cos{\psi_3} \equiv \frac{(\vec{p}_3 \times \vec{p}_{B}) 
              \cdot (\vec{p}_4 \times \vec{p}_5)} 
               { | \vec{p}_3 \times \vec{p}_{B} | \hskip4mm
           | \vec{p}_4 \times \vec{p}_5 | }      \, ,
\end{equation}
where $\vec{p}_{B}$ denotes the direction 
of the proton beam.
$\theta_3$ is the angle of the highest energy jet
with respect to the proton beam direction.
$\psi_3$ is the angle between the plane spanned
by the highest energy jet and the proton beam 
and the plane containing the three jets.
The angle $\psi_3$ indicates whether the third jet
(i.e.\  the lowest energy jet) is radiated within
($\psi_3 \rightarrow 0$ or $\psi_3 \rightarrow \pi$) 
or perpendicular ($\psi_3 \rightarrow \pi/2$) to
the plane containing the highest energy jet and the proton beam. 

The observable $R_{3/2}$, defined by the ratio of the inclusive
three-jet cross section and the inclusive two-jet cross section, 
is of interest especially for quantitative studies, since 
in this ratio both experimental and some theoretical uncertainties 
cancel to a large extent.

This contribution presents measurements of the inclusive three-jet 
cross section in DIS as a function of $Q^2$.
Distributions of three-jet events are measured in the variables 
$\cos \theta_3$ and $\psi_3$.
The ratio $R_{3/2}$ is measured as a function of $Q^2$. 
Measurements of further observables can be found in \cite{bib:paper}.

% **********************************************************************
% **********************************************************************
% **********************************************************************
\section{Results \label{results}}

The analysis is based on data taken 
in positron-proton collisions
with the H1 detector in the years 1995--1997 with 
a positron beam energy of $E_e = 27.5\,{\rm GeV}$ and
a proton beam energy of $E_p = 820\,{\rm GeV}$,
leading to a center-of-mass energy $\sqrt{s} = 300\,{\rm GeV}$.
%
%The experimental procedure for the measurement and the cuts for the event
%selection are described in detail in \cite{bib:paper}.
%
We present results from two event samples which correspond to integrated 
luminosities of ${\cal L}_{\rm int} = 21.1\,{\rm pb}^{-1}$
(low $Q^2$: $5\, < \, Q^2 \,< \,\phantom{0} 100\,{\rm GeV}^2$) and ${\cal L}_{\rm int} = 32.9\,{\rm pb}^{-1}$
(high $Q^2$: $150\, < \, Q^2 \,< \, 5000\,{\rm GeV}^2$), respectively. Jets
are selected in the
Breit frame with $E_T>5\,{\rm GeV}$.
The data are corrected for effects of detector resolution
and acceptance, as well as for inefficiencies of the selection
and for higher order QED effects.

% **********************************************************************
% ***** new Figure 2
\begin{figure}
\vspace{4.5cm}
\begin{center}
\begin{picture}(20,154)
\put(-205,76){\epsfig{file=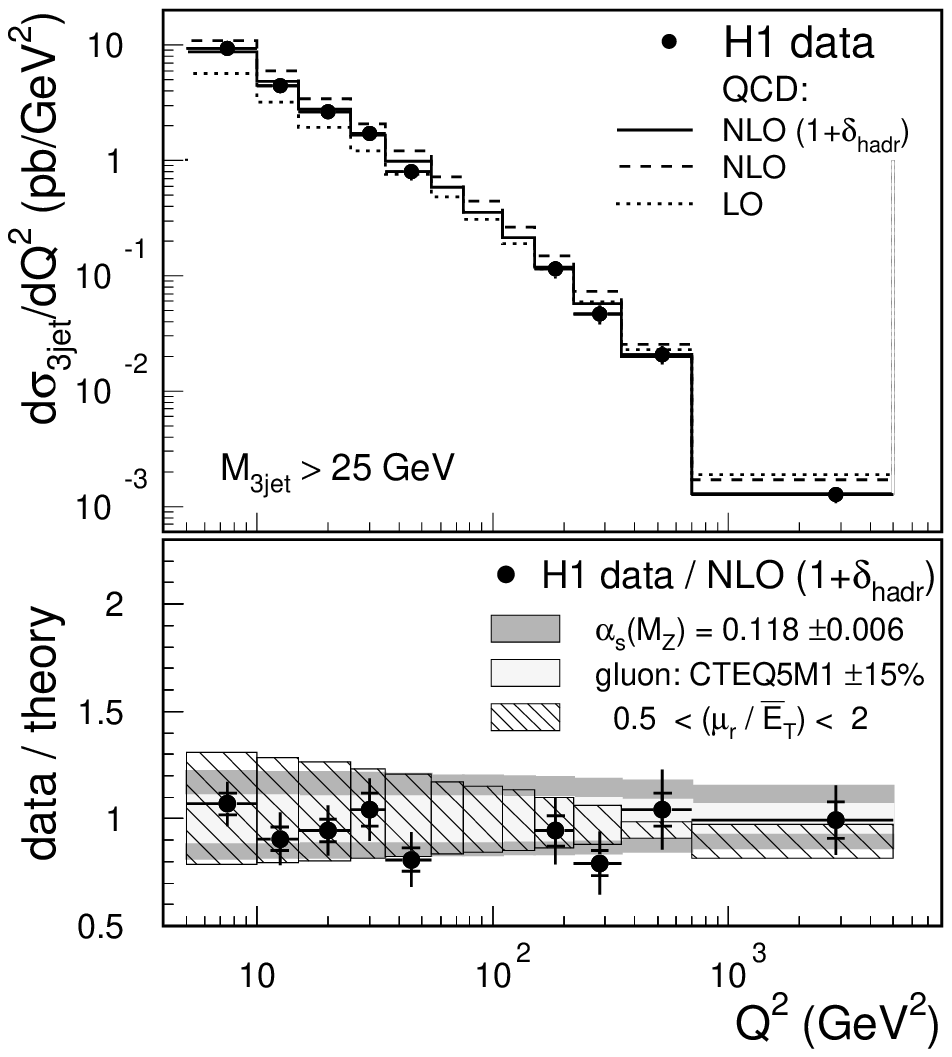,width=7.1cm}} 
\put(-5,76){\epsfig{file=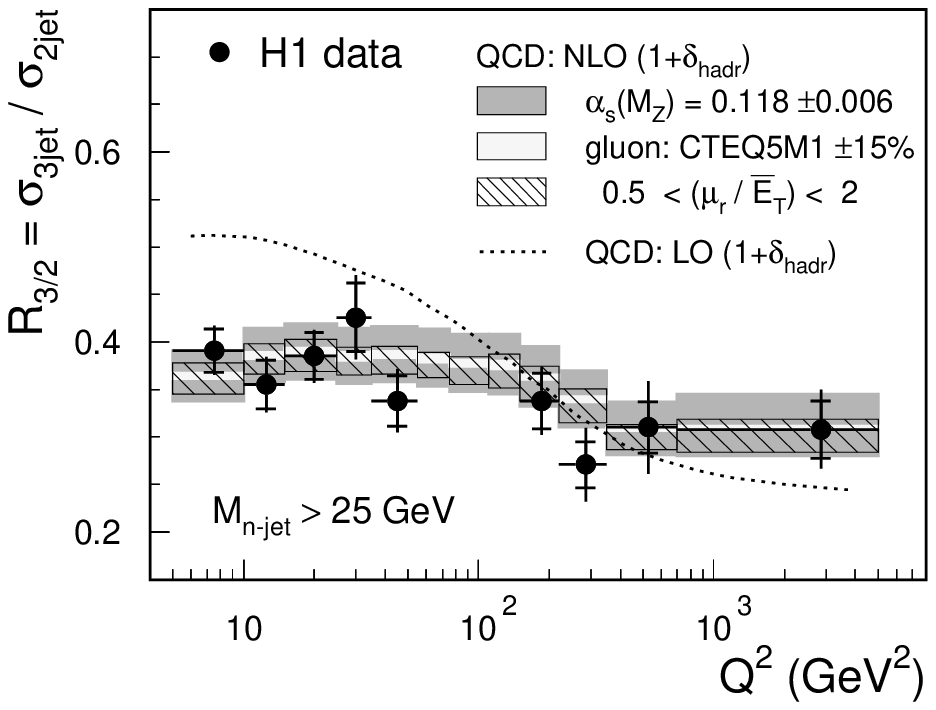,width=7.1cm}}
\put(-160,166){\sf (a)}
\put(42,193){\sf (b)}
\end{picture}
\end{center}
\vskip-10mm
\vspace{-2cm}
\caption{The inclusive three-jet cross section (a) measured as 
a function of the four-momentum transfer squared $Q^2$.
The predictions of perturbative QCD 
in LO (dotted line) and in NLO
with (solid line) and without hadronization corrections 
(dashed line) are compared to the data.
Also shown is the ratio of the measured cross section and the
theoretical prediction, including the effects from variations
of $\alpha_s(M_Z)$, the renormalization scale $\mu_r$ and
the gluon density in the proton.
The ratio $R_{3/2}$ of the inclusive three-jet cross section
to the inclusive dijet cross section (b) is compared to 
the LO (dotted line) and
the NLO calculations (central value of the light band)
including hadronization corrections.
The sensitivity of the NLO calculation to parameter variations
is indicated as in (a).
}
\label{fig:xsectq2}
\end{figure}
% **********************************************************************

The measured three-jet distributions are presented in Figs.~1 and 2.
The inner error bars represent the statistical uncertainties and the 
outer ones the quadratic sum of all uncertainties.
The data are compared to pQCD predictions in LO 
and  
NLO with and without hadronization corrections.
The LO and NLO calculations are carried out in the
$\overline{\rm MS}$-scheme for 
five massless quark flavors using the program 
NLOJET~\cite{nlojet}.
Renormalization and factorization scales ($\mu_r$, $\mu_f$) are set to the 
average transverse energy $\overline{E}_T$ of the three jets 
in the Breit frame.
Hadronization corrections $\delta_{\rm hadr}$ are determined 
using the Monte Carlo event generator
LEPTO~\cite{lepto} as the relative change of an observable before and 
after hadronization.

Fig.~\ref{fig:xsectq2} (a) shows that
over the whole range of $Q^2$ the NLO prediction 
(corrected for hadronization effects) gives a good description 
of the data --- not only at high $Q^2$, where NLO corrections
are small, but also at low $Q^2$, where the NLO prediction
is a factor of two above the LO prediction.
The theoretical prediction is subject to several uncertainties,
the dominant sources being the value of the strong coupling, 
the parton density functions of the proton (especially the gluon density)
and the renormalization scale dependence of the NLO calculation
(see lower part of Fig.~\ref{fig:xsectq2} (a) ).
While for $Q^2 \gtrsim 50\,{\rm GeV}^2$ the variation of $\alpha_s$ 
gives the largest effect, the renormalization scale dependence 
is the dominant source of uncertainty at lower values of $Q^2$,
i.e.\ in the region where NLO corrections are also large.
Over the whole $Q^2$ range the change of the cross section, induced
by the variation of the gluon density is approximately half as large 
as the change induced by the $\alpha_s$ variation.

%The ratio $R_{3/2}$ is shown 
%in Fig.~\ref{fig:xsectq2} (b) as a function of $Q^2$.
%
While the LO calculation predicts a stronger $Q^2$ dependence 
of the ratio $R_{3/2}$ than observed in the data (Fig.~\ref{fig:xsectq2}
(b) ), the NLO calculation
gives a good description of the data over the whole $Q^2$ range.
Uncertainties in the theoretical prediction for
$R_{3/2}(Q^2)$ are investigated in the same way as for 
the three-jet cross section.
The NLO corrections for the three-jet and the dijet cross 
sections are of similar size.
At low $Q^2$ this leads to a smaller NLO correction and 
renormalization scale dependence for the ratio $R_{3/2}$ 
than for the cross section.
Furthermore, when measured in the same region of 
$x_{\rm Bj}$ and $Q^2$, with the same cut on 
the invariant multi-jet mass,
the three-jet and the dijet cross sections 
probe the parton density functions of the proton in the
same range of proton momentum fractions 
$\xi = x_{\rm Bj} (1+ M^2_{\rm n-jet}/Q^2)$.
Since both jet cross sections are dominated by gluon induced
processes, $R_{3/2}$ is almost insensitive to variations
of the gluon density in the proton.
For the central value of $\alpha_s(M_Z) \approx 0.118$ 
used in the calculations,
the theoretical predictions are consistent with the data for
the three-jet cross section and the ratio $R_{3/2}$.

% **********************************************************************
% ***** new Figure 5
\begin{figure}
\vspace{3.5cm}
\begin{center}
\begin{picture}(0,100)
\put(-155, 77){\epsfig{file=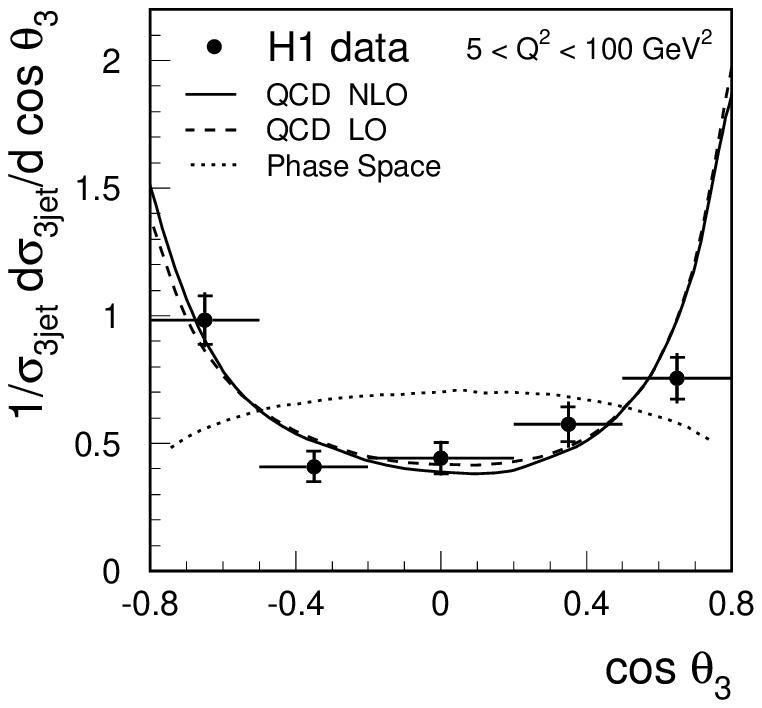,width=5.2cm}} 
\put(   0, 77){\epsfig{file=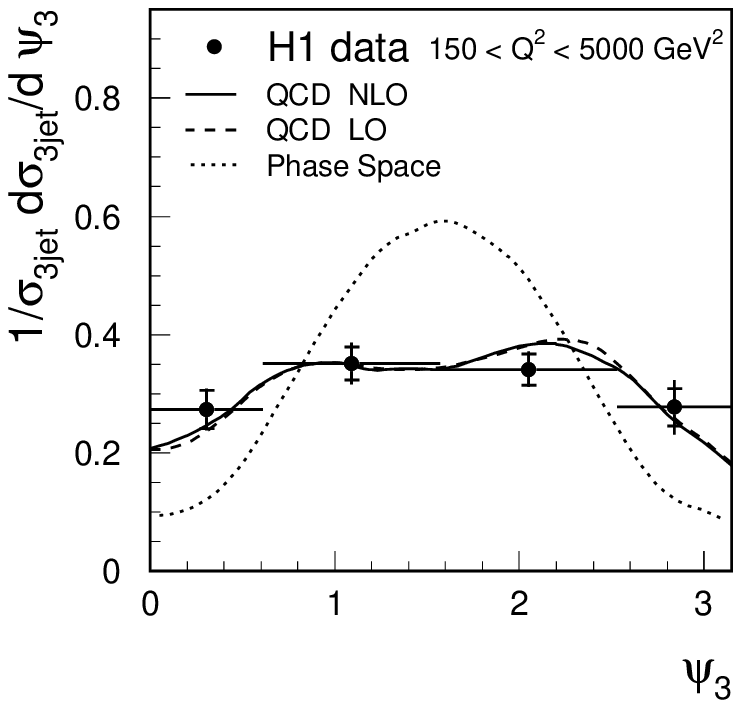,width=5.2cm}}
\put(-112,110){\sf (a)}
\put(48,110){\sf (b)}
\end{picture}
\end{center}
\vspace{-3cm}
\caption{Distributions of $\cos \theta_3$ at low $Q^2$ (a) and $\psi_3$
at high $Q^2$ (b).
The data are compared to the predictions of perturbative QCD 
in NLO (solid line) and in LO (dashed line)
and to a uniform population of the available three-jet phase space (dotted
line).
}
\label{fig:th3psi3lo}
\end{figure}
% **********************************************************************

The normalized distributions of the angular variables 
$\cos{\theta_3}$ and $\psi_3$ are shown in Fig.~\ref{fig:th3psi3lo}.
The $\cos{\theta_3}$ distributions in both $Q^2$ regions (here only the low
$Q^2$ region is shown)
are peaked at the cut value of $\cos{\theta_3} = \pm 0.8$,
corresponding to angles close to the proton and the 
photon direction.
The phase space prediction shows the opposite behavior, 
falling towards  
$\cos{\theta_3} \rightarrow \pm 0.8$.
The QCD calculation in NLO
gives a good description of the effects 
seen in the data. 
This is also the case for the measured $\psi_3$ distributions (here only
the high $Q^2$ region is shown)
which are relatively flat, while the underlying phase space
is peaked at $\psi_3 = \pi /2$.
Although at low $Q^2$ the NLO prediction is almost a factor 
of two higher than the LO prediction,
the shapes of the angular jet distributions are almost unaffected 
by the NLO correction.
QCD calculations, in contrast to the phase space predictions,
indicate that, due to the bremsstrahlung nature of the process,
configurations are preferred in which the plane containing the two lowest
energy jets coincides with the plane spanned by the proton beam and
the highest energy jet. This corresponds to values of $\psi_3
\rightarrow 0$ and $\psi_3 \rightarrow \pi$. The effects of cuts reduce
this preference (Fig.~\ref{fig:th3psi3lo} (b) ).
%
%Differences between the QCD calculation and the phase space 
%prediction indicate that due to the Bremsstrahlung nature
%of the process configurations are 
%preferred in which the plane containing the two lowest energy jets 
%coincides with the plane spanned by the proton beam and the 
%highest energy jet, corresponding to values of 
%$\psi_3 \rightarrow 0$ and $\psi_3 \rightarrow \pi$.
%
These results are in qualitative agreement with 
measurements in $\bar{p}p$ collisions~\cite{sps,tevatron} and
in $\gamma p$ collisions ~\cite{zeus3jets}.
%

% **********************************************************************
% **********************************************************************
% **********************************************************************
\section{Summary}

The inclusive three-jet cross section and the ratio $R_{3/2}$ of the inclusive three-jet and the inclusive
dijet cross section have been measured as a 
function of $Q^2$.
The predictions of perturbative QCD in next-to-leading order
give a good description of the three-jet cross section
and the ratio $R_{3/2}$ over the whole range of $Q^2$
for values of the strong coupling close to the
current world average of $\alpha_s(M_Z) \simeq 0.118$~\cite{bethke2000}.
Angular jet distributions
have been measured in the three-jet center-of-mass frame.
The angular orientation of the three-jet system follows 
the radiation pattern expected from perturbative QCD.
While the angular distributions are not consistent 
with a uniform population of the available phase space,
they are well described by 
the QCD predictions.
This is the first --- successful --- test of {${\cal O}(\alpha_s^3)$}
calculations at HERA.

%\newpage
%\pagebreak[4]

%\section*{Acknowledgments}
%I would like to thank the ``Hadronic Final States and QCD'' Working
%Group for many discussions
%and the opportunity to present these
%results at this very interesting conference.
%My thanks to the organizers.

% **********************************************************************
%
\bibliographystyle{unsrt}
%\bibliography{3jets}

\begin{thebibliography}{10}


\bibitem{h1gluon}
%\cite{Adloff:2001tq}
%\bibitem{Adloff:2001tq}
C.~Adloff {\it et al.}  [H1 Collaboration],
%``Measurement and QCD analysis of jet cross sections in deep-inelastic  positron proton collisions at s**(1/2) of 300-GeV,''
Eur.\ Phys.\ J.\ C {\bf 19} (2001) 289
[hep-ex/0010054].
%%CITATION = HEP-EX 0010054;%%


\bibitem{bib:paper}
C.~Adloff {\it et al.}  [H1 Collaboration],
%``Three-Jet production in Deep-Inelastic Scattering at HERA''
Phys.\ Lett.\ B {\bf 515} (2001) 17
[hep-ex/0106078].


\bibitem{nlojet}
%\cite{Nagy:2001xb}
%\bibitem{Nagy:2001xb}
Z.~Nagy and Z.~Trocsanyi,
%``Multi-jet cross sections in deep inelastic scattering at  next-to-leading order,''
Phys.\ Rev.\ Lett.\  {\bf 87} (2001) 082001 
[hep-ph/0104315].
%%CITATION = HEP-PH 0104315;%%


\bibitem{lepto}
%\cite{Ingelman:1997mq}
%\bibitem{Ingelman:1997mq}
G.~Ingelman, A.~Edin and J.~Rathsman,
%``LEPTO 6.5 - A Monte Carlo Generator for Deep Inelastic Lepton-Nucleon Scattering,''
Comput.\ Phys.\ Commun.\  {\bf 101} (1997) 108
[hep-ph/9605286] (version used: 6.5).
%%CITATION = HEP-PH 9605286;%%

\bibitem{sps}
%\cite{Arnison:1985zm}
%\bibitem{Arnison:1985zm}
G.~Arnison {\it et al.}  [UA1 Collaboration],
%``Comparison Of Three Jet And Two Jet Cross-Sections In P Anti-P Collisions At The Cern Sps P Anti-P Collider,''
Phys.\ Lett.\ B {\bf 158} (1985) 494; \\
%%CITATION = PHLTA,B158,494;%%
%
%\cite{Appel:1986iv}
%\bibitem{Appel:1986iv}
J.~A.~Appel {\it et al.}  [UA2 Collaboration],
%``A Study Of Three Jet Events At The Cern Anti-P P Collider,''
Z.\ Phys.\ C {\bf 30} (1986) 341.
%%CITATION = ZEPYA,C30,341;%%


\bibitem{tevatron}
%
%\cite{Abachi:1996zv}
%\bibitem{Abachi:1996zv}
S.~Abachi {\it et al.}  [D0 Collaboration],
%``Studies of Topological Distributions of the Three- and Four-Jet Events in $\pp$ Collisions at $\sqrt{s}=1800$ GeV with the D\O\ Detector,''
Phys.\ Rev.\ D {\bf 53} (1996) 6000
[hep-ex/9509005]; \\
%%CITATION = HEP-EX 9509005;%%
%
%\cite{Abe:1992ui}
%\bibitem{Abe:1992ui}
F.~Abe {\it et al.}  [CDF Collaboration],
%``The Topology of three jet events in anti-p p collisions at S**(1/2) = 1.8-TeV,''
Phys.\ Rev.\ D {\bf 45} (1992) 1448.
%%CITATION = PHRVA,D45,1448;%%


\bibitem{zeus3jets}
%\cite{Breitweg:1998uv}
%\bibitem{Breitweg:1998uv}
J.~Breitweg {\it et al.}  [ZEUS Collaboration],
%``Measurement of three-jet distributions in photoproduction at HERA,''
Phys.\ Lett.\ B {\bf 443} (1998) 394
[hep-ex/9810046].
%%CITATION = HEP-EX 9810046;%%


\bibitem{bethke2000}
%
%\cite{Groom:2000in}
%\bibitem{Groom:2000in}
D.~E.~Groom {\it et al.}  [Particle Data Group Collaboration],
%``Review of particle physics,''
Eur.\ Phys.\ J.\ C {\bf 15} (2000) 1.
%%CITATION = EPHJA,C15,1;%%




\end{thebibliography}

\clearpage

\end{document}